%% file: main.tex
\documentclass[10pt]{article}
\usepackage{graphicx,amssymb, amstext, amsmath, epstopdf, booktabs, verbatim, gensymb, geometry, appendix, lmodern,bchart,pgfplots,url,hyperref}
\usepackage{chngcntr}
\usepackage{color, colortbl}
\usepackage{appendix}
\usepackage{wrapfig}
\usepackage{lscape}
\usepackage{rotating}
\usepackage{epstopdf}

\hypersetup{
    colorlinks=true,
    linkcolor=blue,
    filecolor=blue,      
    urlcolor=blue,
    bookmarks=true,
    citecolor=blue,
    pdfpagemode=FullScreen,
    }

\counterwithout{figure}{section}
\counterwithout{table}{section}

\pgfplotsset{width=7cm,compat=1.8}

\newcommand*\Title{KOALA Project Report 3.5}
\newcommand*\cpiType{--- KOALA Project Report 3.5}
\newcommand*\Date{April 2019}

\title{What concerns do Chinese parents have about their children's digital adoption and how to better support them ?}

\author{Ge Wang \\ Jun Zhao \\ Nigel Shadbolt}
\date{\today}

\usepackage{cpi} 

\begin{document}

\begin{titlepage}
\maketitle
\end{titlepage}

\linespread{1.15} 

\begin{executive}

Digital devices are widely used by children, children nowadays are spending more time online than with other media sources, such as watching television or playing offline video games. In the UK, 44\% of children aged five to ten have been provided with their own tablets, with this percentage increasing annually \cite{ofcom2017}, while in the US, ownership of tablets by children in this age group grew fivefold between 2011 and 2013 \cite{commonsense2013}. 

Our previous research found that UK children and parents need better support in dealing with online privacy risks. Interestingly, very few research were done on Chinese children and parents. According to CNNIC (China Internet Network Information Center), 89.5\% of Chinese children aged 6-11 have access to the Internet - watching videos online and online gaming were reported to be the main reason for children going online. Most used devices by teenagers includes smartphones (92\%), PCs (48.7\%), smart TVs (47.7\%) and tablets (37.4\%). The same report shows that Chinese parents showed increasing concerns on topics including online baiting, inappropriate contents and cyber-bulling \cite{cnnic-teen18}. However, less awareness has been drawn on children's online privacy. 

In this report, we present findings from our online survey of 593 Chinese parents with children aged 6-10 in February and March 2019. Our survey particularly focused on understanding Chinese' parents' awareness and management of their children's online privacy risks. Participant families were mainly from Shanghai, the largest cosmopolitan city in China, as well as surrounding areas. The goal of the survey was to examine the current adoption pattern of digital devices by Chinese families with young children, the concerns Chinese parents have about their children's online activities and the current practices they use for safeguarding their children online. 

\frame{
\textbf{Four key findings}
\begin{enumerate}
    \item Digital devices were widely adopted in Chinese families. Parents showed some privacy concerns in general, however, their primary concern were still on the content their children might have access to and screen time control, rather than what personal information that might be collected from their children.
    
    \item Parents' levels of concerns can be influenced by their own digital experiences. Parents with more digital experiences can have a higher level of concerns of their children's privacy online.
    
    \item Online short-video platforms played an important role in Chinese young children's daily life, however, many of these apps are not always appropriate for children's age. Online learning was reported to be another major reason for children being online. We found that schools and teachers played an important role in children's choices of apps and this is largely different from the UK children.
    
    \item Most parents used a range of means to safeguard their children online, however mostly through restrictive approaches. Only a small proportion of them (26.6\%) regularly discussed privacy issues with their children and very few of them had sufficient awareness of the potential risks (only 10\% think there exists noticeable privacy risks in their children's daily online activities).
    This shows that parents would benefit from support in tools and indicates a need for resources to help parents safeguard their children online. 

\end{enumerate}
}

These findings imply that we need to continue presenting specific guidance to parents in order to support their choice of digital content for their young children. Further, we need to look more deeply into the roles schools are taking in children's online activities, how can we support schools and teachers when they are making recommendations to parents and children. Finally, we need look more deeply into whether current guidance on apps' age appropriateness is sufficiently useful for parents, and how parents can be better supported to mediate the choice of digital content by involving their children. Most of implications are not unique to Chinese parents; however, Chinese families face more influence from schools in terms of their choice of technologies, which are intended for education and learning purposes.

\end{executive}

\input{introduction}

\input{method}

\input{discussion}

\input{conclusion}

\section{Acknowledgement} 

We thank all the schools, families and children who have contributed their time and knowledge to our study. Without their support, this study would not have been successful!

This research is supported by KOALA (http://SOCIAM.org/project/koala): Kids Online Anonymity \& Lifelong Autonomy, funded by EPSRC Impact Acceleration Account Award, under the grant number of EP/R511742/1.

\section{Contact} 
\begin{itemize}
    \item Jun Zhao: jun.zhao@cs.ox.ac.uk
    \item Nigel Shadbolt: nigel.shadbolt@cs.ox.ac.uk
\end{itemize}

\bibliographystyle{SIGCHI-Reference-Format}
\bibliography{koala,chi-2017,kids-studies}

\end{document}

%% file: introduction.tex
\section{Introduction}

Digital devices are widely adopted by children nowadays and they are largely used to complement education and provide entertainment at home and school. According to the 2016 "Children and Parents' Media Use and Attitudes'' report (Ofcom)~\cite{ofcom2016}, children aged 5 to 15 years are, for the first time, spending more time online than watching TV. Among the many kinds of devices now connected to the Internet, mobile devices and tablets have become the primary means by which children go online~\cite{livingstone2017children}. Our previous findings have shown that parents within the UK and other western countries are generally concerned about their children's digital adoptions and their online privacy. However, the major focus is on the content their children might be seeing. Most parents use a range of technical restrictions to safeguard their children, and it is largely agreed that parents these days need better support for safeguarding their children's online safety, and for making informed choices about digital content consumed by their children~\cite{livingstone2018parentsa}.

We want to extend this research beyond western countries, and this report will be an addendum to our existing findings. According to CASS's Bluebook of Teenagers, the proportion of Chinese children under 10 who use the Internet - which was only 56\% in 2010 - reached 68\% in 2018~\cite{cass2018}. This rapidly growing adoption of digital devices has raised new privacy risks and challenges for families and parenting~\cite{livingstone2014eu}. Existing research on EU/western families has found that:

\begin{itemize}
    \item Digital adoption is deeply rooted in family lives~\cite{DBLP:journals/corr/abs-1809-10841}, however, parents sometimes find it quite hard to keep up with the rapid developing technologies. Digital skills are not universal amongst parents and children, especially those related to online privacy issues~\cite{livingstone2018parentsc}. Parents need better support to manage their struggle of mediating their children's choice of apps~\cite{zaman2016parental}. This is important for not only fostering a positive parenting experience but also for transmitting the essential knowledge and skills to young children, who are at the frontier of risks~\cite{DBLP:journals/corr/abs-1809-10841}.
    \item Parents often think their children are too young to understand privacy risks online and delay these conversations with their children. This leaves young children mostly under their parents' protections and could potentially lead to lacking of the essential skills to fend themselves or seek help when needed~\cite{DBLP:journals/corr/abs-1809-10944}.
    \item Children care about their privacy online, and are sensitive to, who might access their sensitive information (e.g. real names, age, location etc), and could apply a range of techniques to safeguard this space. However, they still need help to fully understand online privacy risks and when they struggle, they often seek help from their parents~\cite{10.1145/3290605.3300336}. However, parents do not necessarily fully understand the risks themselves~\cite{DBLP:journals/corr/abs-1809-10944}. 
\end{itemize}

It is largely agreed that parents these days need better support for safeguarding their children's online safety, and for making informed choices about digital content consumed by their children~\cite{livingstone2018parentsa}. However, rather limited research has been done specifically on Chinese families, and very few research has looked at what concerns Chinese parents have in terms of their children's digital adoption. This report provides a first-step understanding about what Chinese parents currently struggle with, so that we can design and develop the kind of support parents most need, and encourage a positive co-learning process for both parents and their children.

\section{Background of Internet Development in China}

The uptake of Internet in China is rapidly changing. Not only the number of population in China that adopts the technologies continues to grow in general --- nearly 60\% of the overall population now going online, the online proportion in the rural area also continues to increase~\cite{cnnic43}. 

Mobile platforms are the dominant means for Chinese Internet users to go online, with 98.6\% of the online population going online via mobile phones~\cite{cnnic43}. Mobile shopping, payment, online video and short video platforms are among the most popular services. 4.5 billion mobile apps can be found in the Chinese app market, with nearly 60\% of them developed by Chinese companies and game apps taking up over 30\% of the market~\cite{cnnic43}. The CCNIC survey~\cite{cnnic43} shows that on average, Chinese Internet users would spend about 27.6 hours per week online, which is about 4 hours a day. Most of their online activities are centred around communications or online entertainment (like video, music, literature reading etc).

At the same time, the proportion of Chinese children going online also increases. China is now home to 169 million Internet users under the age of 18, 89.5\% of children under 13s have been reported to have access to the internet ~\cite{cnnic-teen18}. While mobile phones were still the major way for teenagers going online (92\%), tablets (37.4\%) and smart TVs (46.7\%) were among the devices most frequently used and have been used more by teenagers than the other age groups. The 2018 CNNIC report ~\cite{cnnic-teen18} shows that 77.6\% of those under 18s have their own devices. Online learning, online video and gaming are among the most popular services. The emerging online short video platforms (users make videos themselves) are widely favored, with nearly 41\% of usage ratio.

Alongside with the rapid increase in online adoption of Chinese children, there have been growing concerns. For those under 18s, 30.3\% have had exposure to inappropriate contents and 15.6\% had experienced online bullying ~\cite{cnnic-teen18}. However, those privacy-related risks were not looked at or discussed. Recently, with the disclosure of Facebook's privacy incidents and the remarks made by a renowned tech giant - "users are willing to trade privacy for convenience", online privacy concerns has been flagged up once again. As the generation growing up at the frontier of IOT, children's daily activities are constantly shifting from 'offline' to 'online'. Both the amount of information, and the value of them has been continuously increasing, and there has been increasing risk of children's privacy being compromised or improperly exploited.

According to the "Android Application Security White Paper" released by Tencent Cohen Lab, the examine of the 1404 most downloaded apps in 2018 found that 92\% of Android apps requested excessive core privacy permissions \cite{kennlab}, the private data collected includes personal information such as location information, address book information, and mobile phone number. For example, some school management apps required parents to submit the student's national school registration number, school class, ID number, location, parent address, contact number and other details. This has also led parents to worry about potential privacy breaches of children and family, and increased concerns on the ownership of these personal information.

This has prompted us to look into the online activities of Chinese children. The goal of our study is to investigate the digital adoption patterns by Chinese children, from device usage to online activities, and to discuss what concerns do Chinese parents have. How can they be better supported, what mechanisms would they prefer to safeguard their children online, and do they have awareness and needs in terms of their children's online privacy risks.

%% file: method.tex
\section{Methodology}
Our report is based on an online survey conducted in March 2019. Participants were recruited through local primary schools in Shanghai area, and the survey was hosted by the Online Surveys platform ($https://www.onlinesurveys.ac.uk$) in the UK, provided by the Jisc, a UK-based non-profit company. 

In our survey, participants were required to be parents of, or guardians to, at least one child aged between 6 and 10, who has regular access to a tablet computer or smartphone. Participants were largely recruited from the same residential area, however, of varied levels of educational levels.

Questions in the survey were based on our previous online surveys with UK parents~\cite{DBLP:journals/corr/abs-1809-10841}, and related literature. The goal of the survey was to extend our samples from within UK to Chinese parents, and understand the following:

\begin{itemize}
    \item Are digital devices widely adopted by Chinese children, what is the role of technologies in their daily life and what are the digital devices they mostly interact with?
    \item What concerns do Chinese parents have on their children's digital adoption and the related privacy risks? How do they see the balance between opportunities and risks?
    \item What are parents' perceptions of children's online privacy, what are their current barriers? Are they aware of algorithm-based processing of their children's data? What strategies do parents develop to mitigate the potential risks, and who or where do they go for advice? How can they be better supported?
\end{itemize}

The online survey received 593 responses. Among them, the average age of parents was 35.4, and 73.4\% was female. All of our respondents were Chinese residents, most of them (92\%) resided in Shanghai and the remaining 8\% resided in suburb areas around Shanghai. Most of our respondents were on full-time employment (60.5\%), more than half of them (52.8\%) claimed that they have not done job related to, or do not have extensive knowledge of digital technologies. Around 31.9\% of our respondents had degree-level education and above, 28.7\% had diploma-level education, while the remaining 39.4\% had high school level education and below. The average age of the child selected for completing the survey was 8.49.

Details about the methodology, including the survey design and additional data tables can be made available upon request.

%% file: discussion.tex
\section{Key Findings}
\subsection{Mobile devices are dominant in Chinese families and children spent more time online if their parents do}

\begin{figure}[h]
\centering
\includegraphics[width=0.9\textwidth]{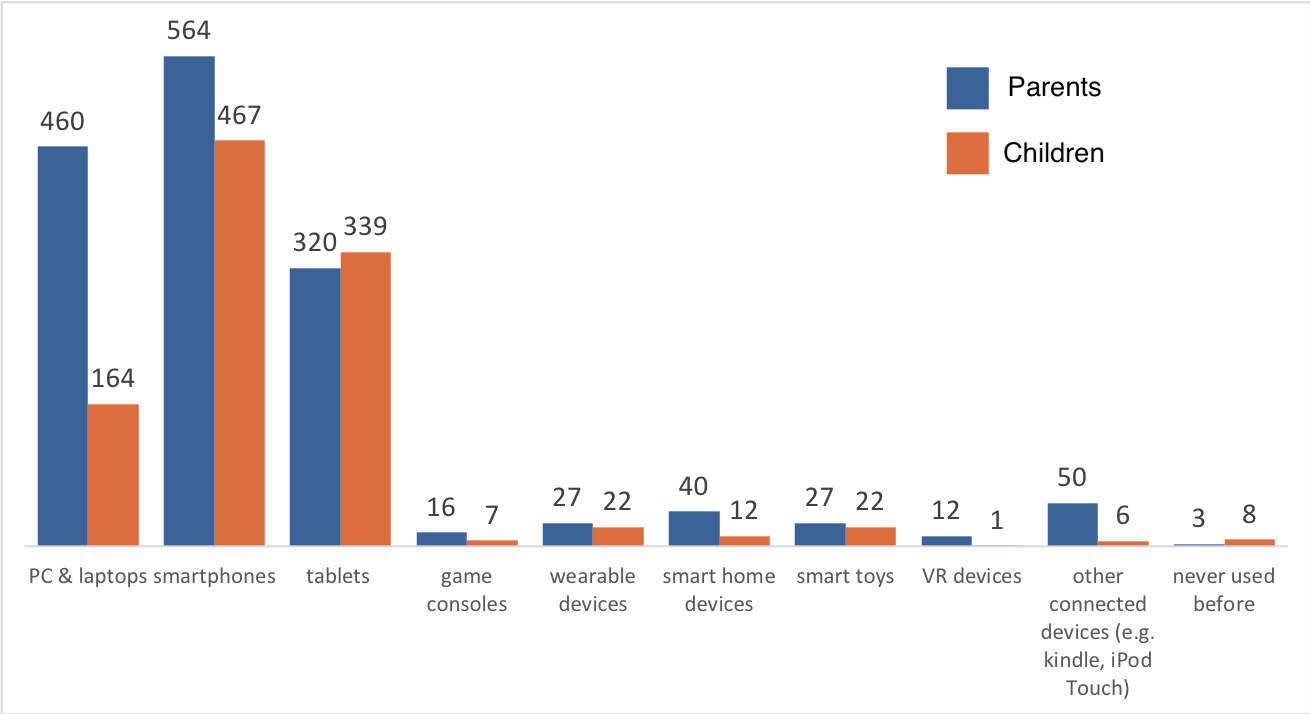}
\caption{Devices used by parents/children in the past month}
\label{fig:q7_11}
\end{figure}

\begin{figure}[h]
\centering
\includegraphics[width=0.8\textwidth]{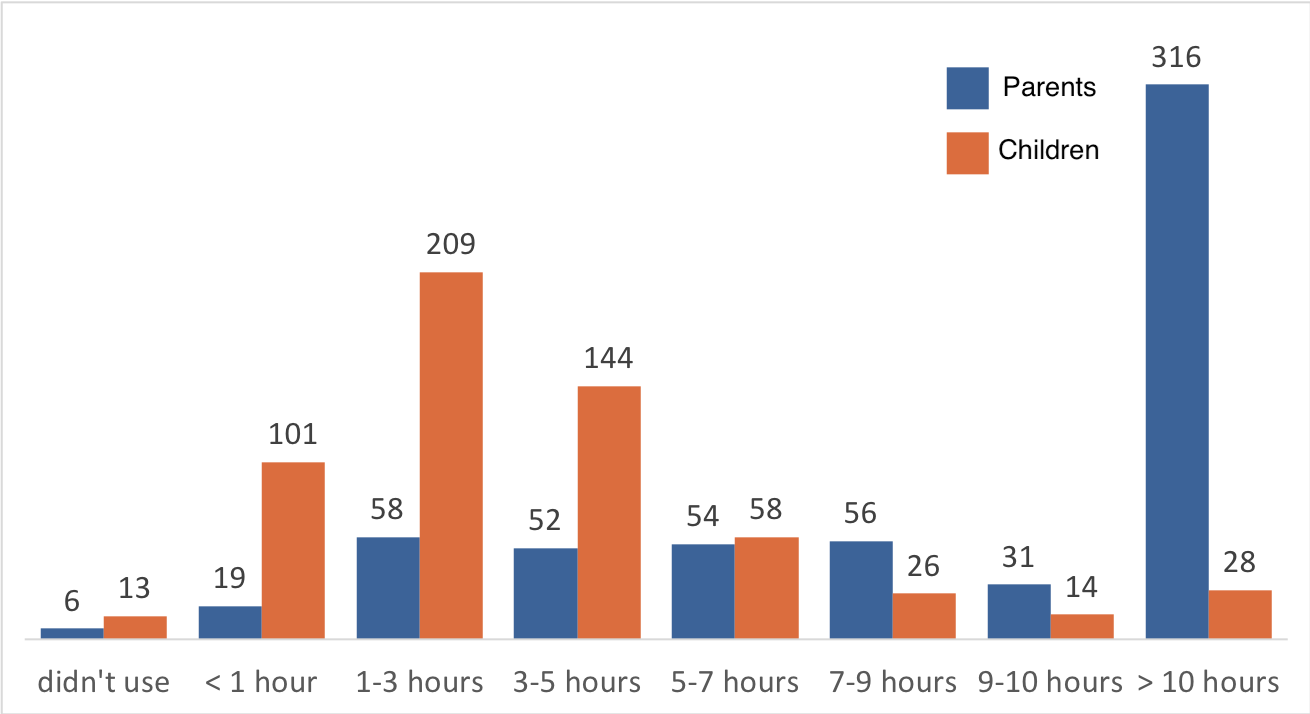}
\caption{How many hours do parents/children spend online last week}
\label{fig:q8_12}
\end{figure}

\begin{figure}[h]
\centering
\includegraphics[width=1\textwidth]{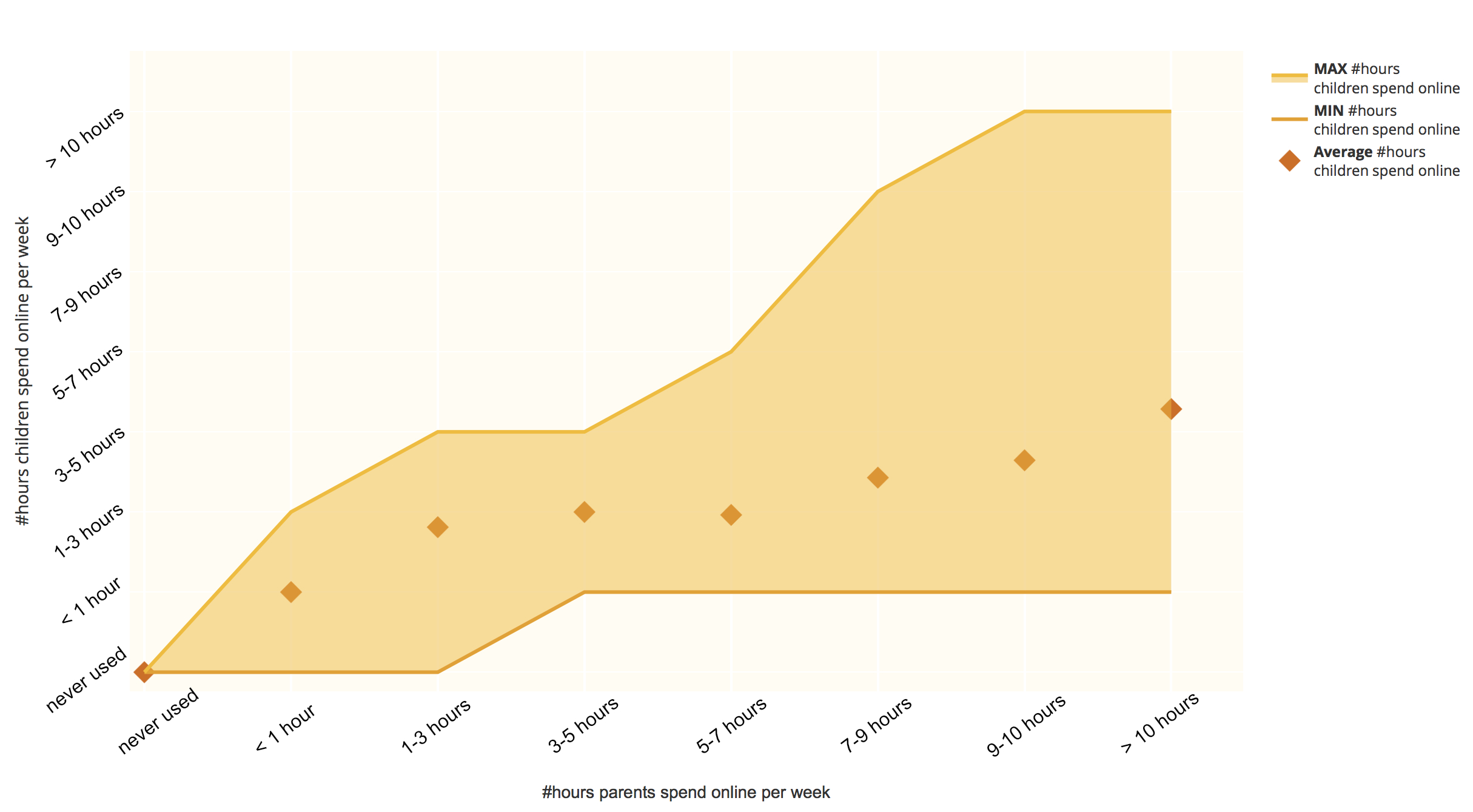}
\caption{As parents spend more time online, children spend more time online.}
\label{fig:onlinetimecorr}
\end{figure}

In the survey, parents were asked to report their digital adoption patterns (\textbf{Q7-8}) and that of their children (\textbf{Q11-12}). Computers, smartphones and tablets were reported to be the most used devices by parents and children to go online. As for the children, smartphones were used most frequently (78.8\%), followed by tablets (57.2\%). While the majority of parents go online for more than 10 hours per week (53.4\%), the children's online hours per week were most frequently claimed to be around 1-3 hours (35.2\%) and 3-5 hours (24.3\%). Children spent more time online if their parents do (Figure ~\ref{fig:q7_11}, ~\ref{fig:q8_12}, ~\ref{fig:onlinetimecorr} )

\subsection{Children have certain level of autonomy when choosing apps, while still being largely under parents' control.}
\begin{figure}[h]
\centering
\includegraphics[width=0.9\textwidth]{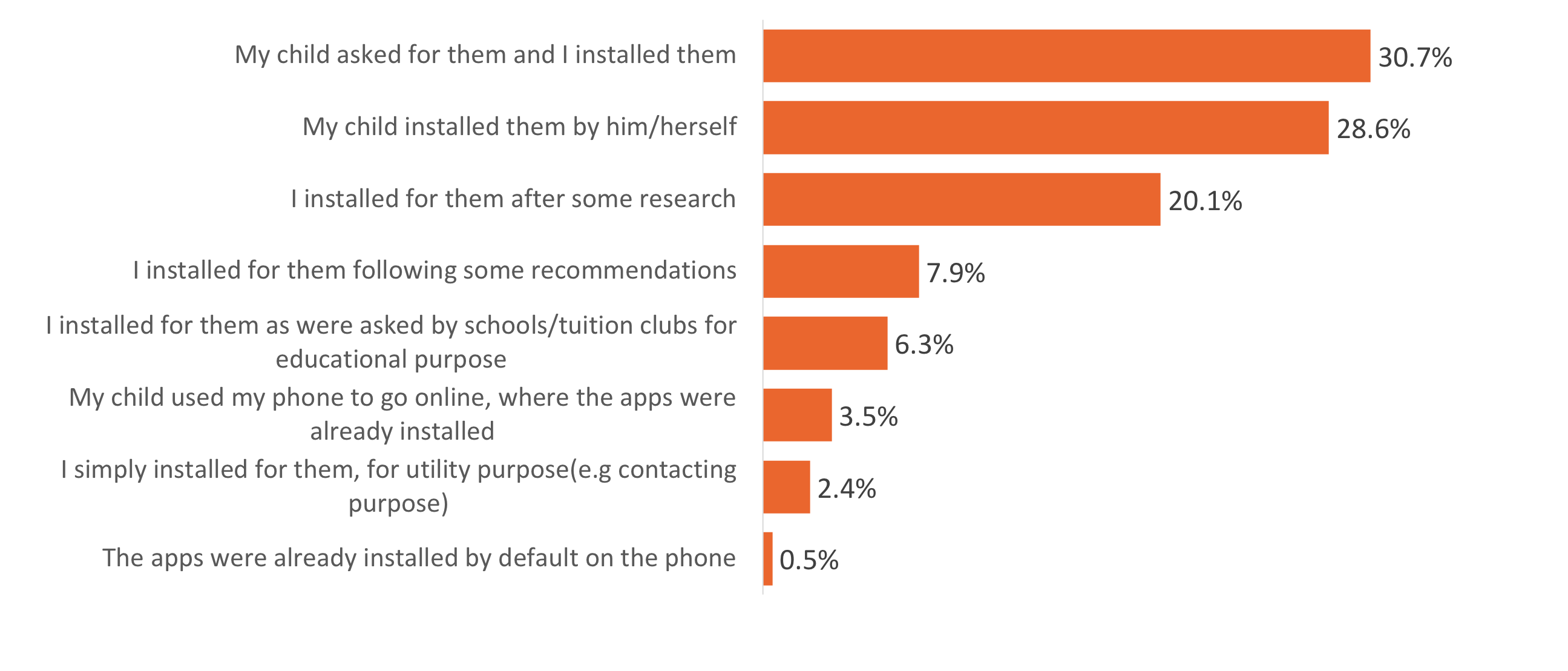}
\caption{How did your child get these apps installed on the device? (Q16)}
\label{fig:q16}
\end{figure}

In the survey, parents were asked to report two of their child's most used mobile apps (\textbf{Q15}) and how they were installed (\textbf{Q16}) (Figure ~\ref{fig:q16}).

\begin{itemize}
    \item 30.7\% of parent respondents said that they had installed the apps after the child had asked for it, 28\% of them said that they had installed the apps after some research or following some recommendation. 
    \item A fair amount (28.6\%) reported that the child installed the apps by themselves.
    \item Although not given as an option in the survey, a considerable amount (6.3\%) of parent respondents reported that they had installed the apps as were required by schools/tuition clubs for educational purpose, which correlates with their claims of their child's main purpose online was for educational needs.
\end{itemize}

In comparison to our previous survey to the UK parents~\cite{DBLP:journals/corr/abs-1809-10841}, a much bigger proportion of Chinese children are reported to install apps by themselves, while UK parents reported only 19\% of their children are installing apps by themselves. At the same time, a much larger number of UK parents (71\%) reported that they installed the apps for their children than Chinese parents (30.7\%). This may be associated with the relevant lower level of digital concerns of the Chinese parents who participated in the survey. While online privacy being the top barrier for parents' internet use (12\% of all parents as reported in Livingstone's report \cite{livingstone2018parentsc}), 40.7\% of Chinese parents think there are very few privacy risks for their children or never thought about this problem, and only 10\% considered there to be privacy risks worth noticing.

\subsection{Children's most used apps are not always appropriate for their age}
\begin{figure}[h]
\centering
\includegraphics[width=0.8\textwidth]{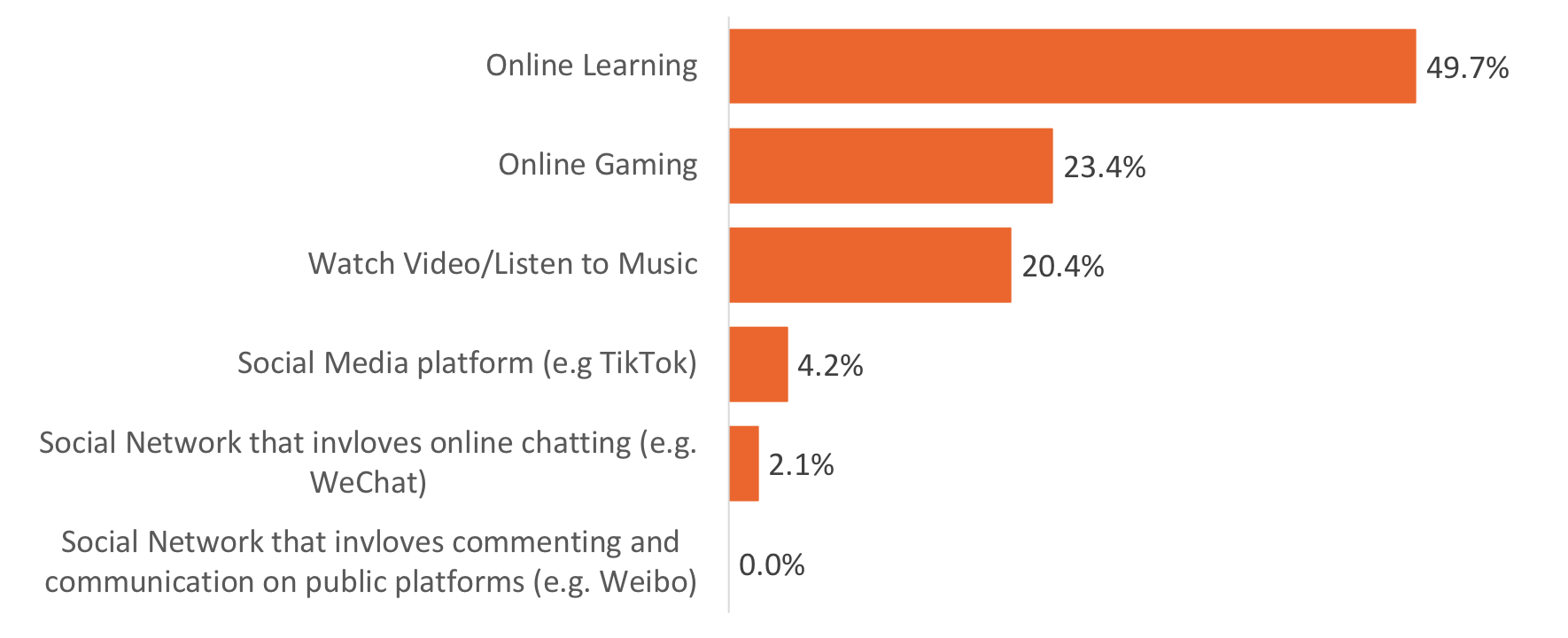}
\caption{In general, which one do you think is the primary purpose of your child's online access? (Q13)}
\label{fig:q13}
\end{figure}

\begin{figure}[h]
\centering
\includegraphics[width=1\textwidth]{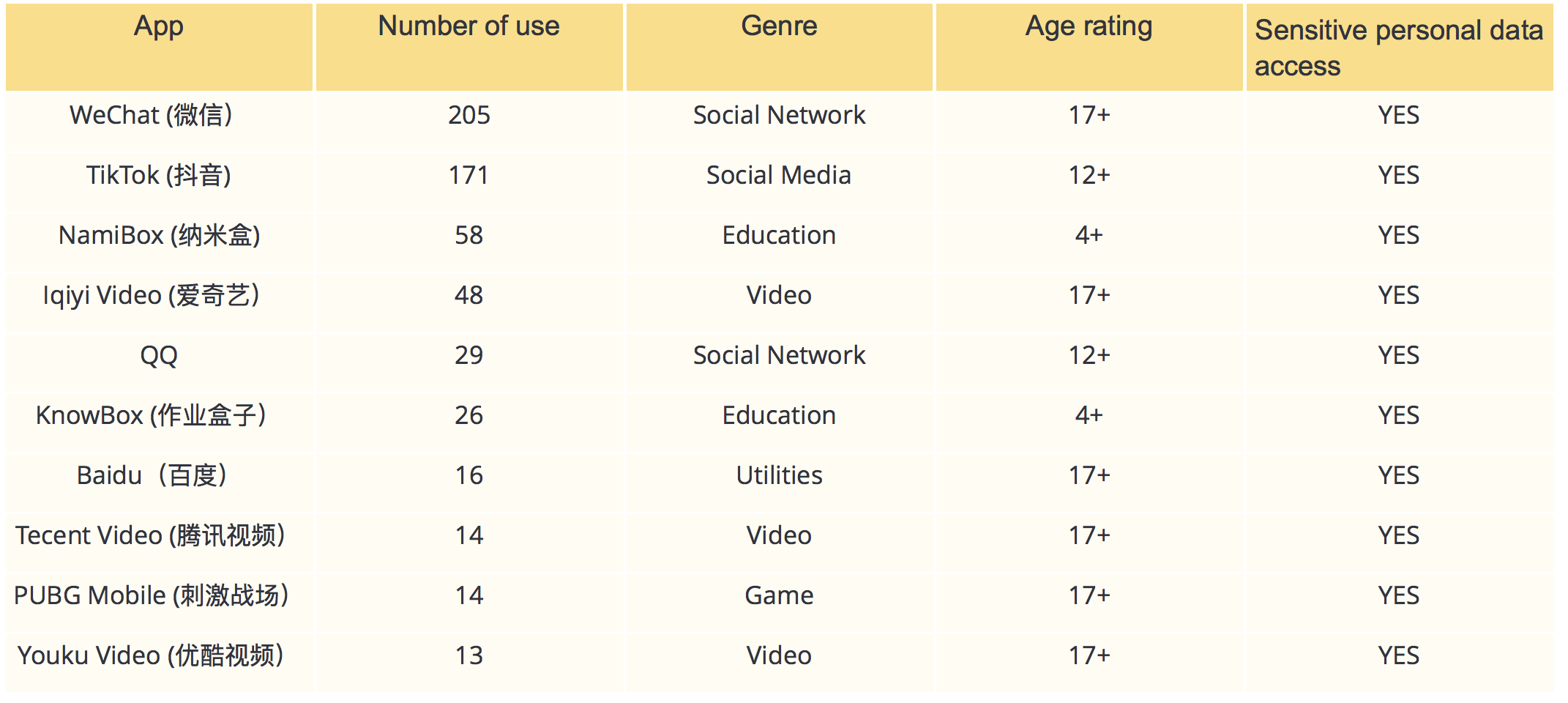}
\caption{10 Apps that were used most frequently as mentioned in the survey (Q15)}
\label{fig:top10app}
\end{figure}

In the survey, parents were asked about what are the main purpose for their children going online (\textbf{Q13})(Figure ~\ref{fig:q13}) and their children's most used apps (\textbf{Q15})(Figure ~\ref{fig:top10app}).
\begin{itemize}
    \item 49.7\% of parent respondents said that their children's main purpose of online access is online learning, only very few reported that the children's main purpose online was using social media platforms (4.2\%) or social networking/online chatting platforms (2.1\%).
    \item However, when asked about their children's most used apps on their devices, we started to see conflicts between what parents have reported. WeChat - a social networking/online chatting platform, was the app most frequently mentioned (34.6\% of all respondents) by the parents as one of their child's two most used apps, followed by TikTok - a social media platform (28.8\%) and NamiBox - an online learning platform (9.8\%).
\end{itemize}

Figure ~\ref{fig:top10app} shows the top 10 apps that were reported to be most used by children in the survey responses:
\begin{itemize}
    \item \textbf{8 out of 10 had an age rating inappropriate \footnote{The information about each app is retrieved from Apple App Store, accessed in April 2019.} for young children aged 6-10}. 
    \item \textbf{All apps required access to sensitive personal data \footnote{Sensitive personal information refers to the type of information that might reveal a person's identity, such as their unique device ID, location or contact details.}}, such as contact details on the device, location information or unique identification of the device.
\end{itemize}

Overall, in comparison to UK children~\cite{DBLP:journals/corr/abs-1809-10841}, Chinese children spend more time on social media platform and online videos than games. This may be related to the wide adoption of social media among Chinese Internet users and how wechat is becoming a multipurpose platform for communication as well as entertainment and digital payment~\cite{cnnic43}.

It is worth noting that, both WeChat and TikTok shared very sensitive information about users (including user's real name, user alias, mobile phone number, password, gender, IP address and credit card information) among third parties and related group companies of Tencent/Music-ly. Although both of the above apps have set their age limit to 12+, it is shown that this does not prevent 6-10 year-olds from using them, in fact, there have been plenty of young tween users. In Feburary 2019, The Federal Trade Commission announced a \$5.7 million settlement over TikTok, over accusations that the company's app illegally collected personal information about children. The agency found a large percentage of the app's users were under 13 and revealed sensitive personal information including their email addresses, names and schools, which was a violation to Children's Online Privacy Protection Act (COPPA) ~\cite{cecilia2019}. 

\subsection{Parents expressed some level of privacy concerns, but still cared more about content and screen time}

Nearly half of Chinese parents expressed concerns about online privacy in relation to their child's interaction with table devices (\textbf{Q22}), with 47.5\% saying that they thought there were `some, but acceptable' privacy concerns around their child, and 10\% thought they had `a lot' of privacy concerns. However, 32.4\% of parents thought they had no or `very little' privacy concerns, while the remaining 8.3\% had never thought about this problem. 62\% of parents reported that they had never discussed privacy issues with their child, or very rarely (\textbf{Q21}). This shows a much lower level of privacy concerns than UK parents, who reported privacy is their main barrier of going online \cite{livingstone2018parentsc}.

\textbf{Content} and \textbf{Screen time} were  the primary topics Chinese parents would care more about, and were the dominant factors for Chinese parents to choose apps to install or decide to uninstall an app for their young children.

\begin{figure}[h]
\centering
\includegraphics[width=0.9\textwidth]{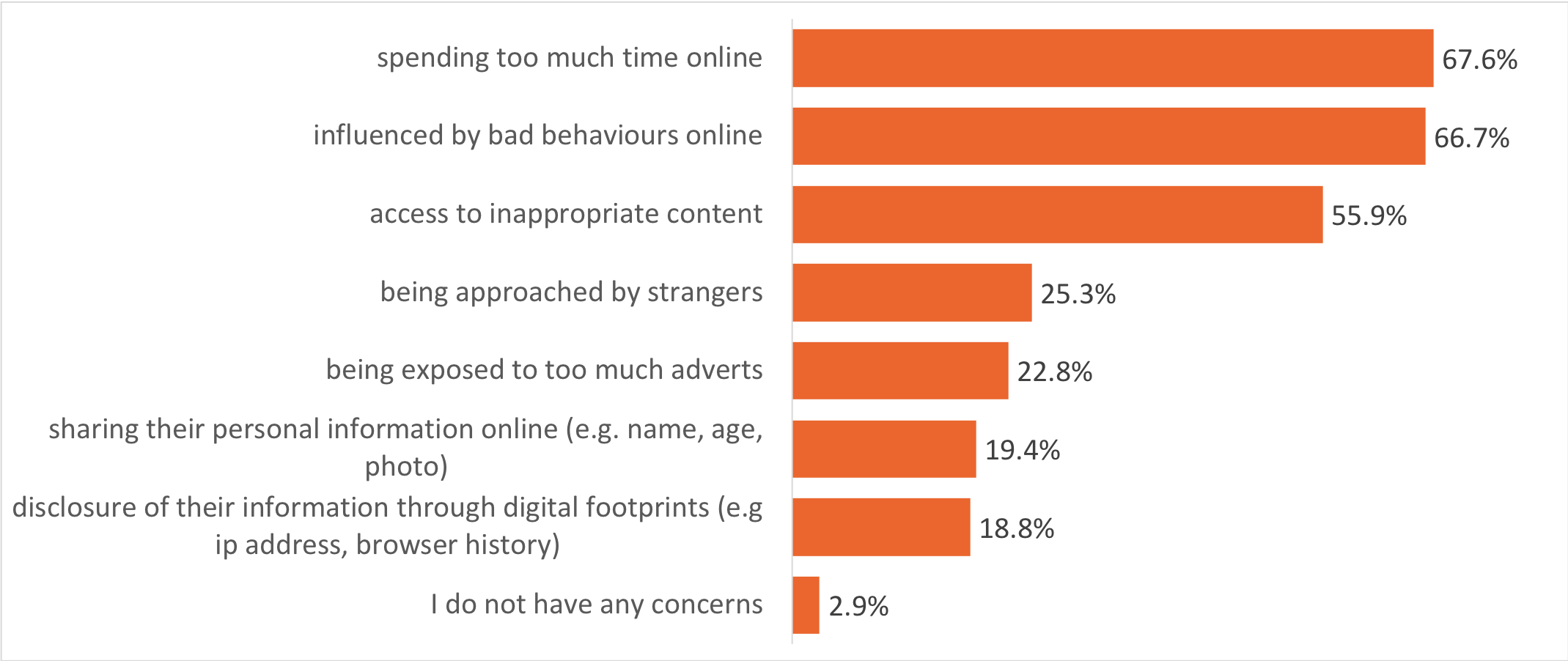}
\caption{When your child is going online, which of the above issues will you be concerned with? (Q14)}
\label{fig:q14}

\centering
\includegraphics[width=0.8\textwidth]{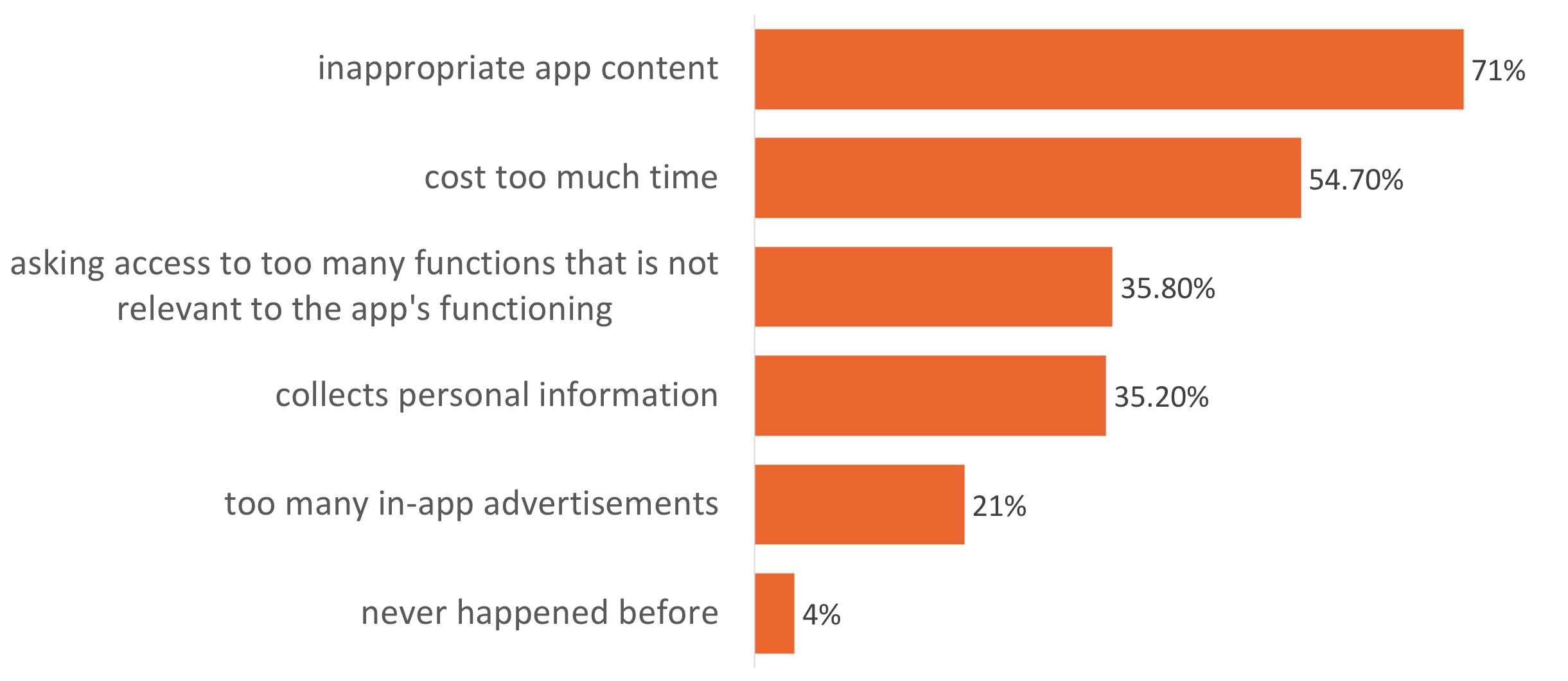}
\caption{Think about the case when you uninstall your child's app, which of the above issues will you be concerned with? (Q18)}
\label{fig:q18}
\end{figure}

\begin{itemize}
    \item 67.6\% of parent respondents were worried about their children might be spending too much time online; 66.7\% of them were concerned with influence by bad behaviours online, which is related to having access to inappropriate content (55.9\%, Figure  ~\ref{fig:q14})
    \item Almost all parents had refused to install or had uninstalled an app (98\%) in the past. The primary reason mentioned by 71\% of parents was that the app content was inappropriate. 54.7\% of parents said that they had done so because the app was too time-consuming. A smaller number of parents (35.8\%) mentioned the reason of the app asking access to things not necessary for the app to function such as camera or location, or specifically that the app was accessing too much information about their child (35.2\%). A few of them mentioned their concerns about in-app advertisements (21\%, Figure ~\ref{fig:q18})
\end{itemize}

In comparison to UK parents, Chinese parents in our survey were much more likely to uninstall their children's apps than UK parents, only 78\% of them reported having removed apps from their children's devices~\cite{DBLP:journals/corr/abs-1809-10841}. Chinese parents also showed a higher level of concerns of screentime and content appropriateness than UK parents. This may be related to the authoritarian parenting style more often observed in Chinese families. 

\subsection{More digitally experienced parents expressed more privacy concerns}
\begin{figure}[h]
\centering
\includegraphics[width=0.8\textwidth]{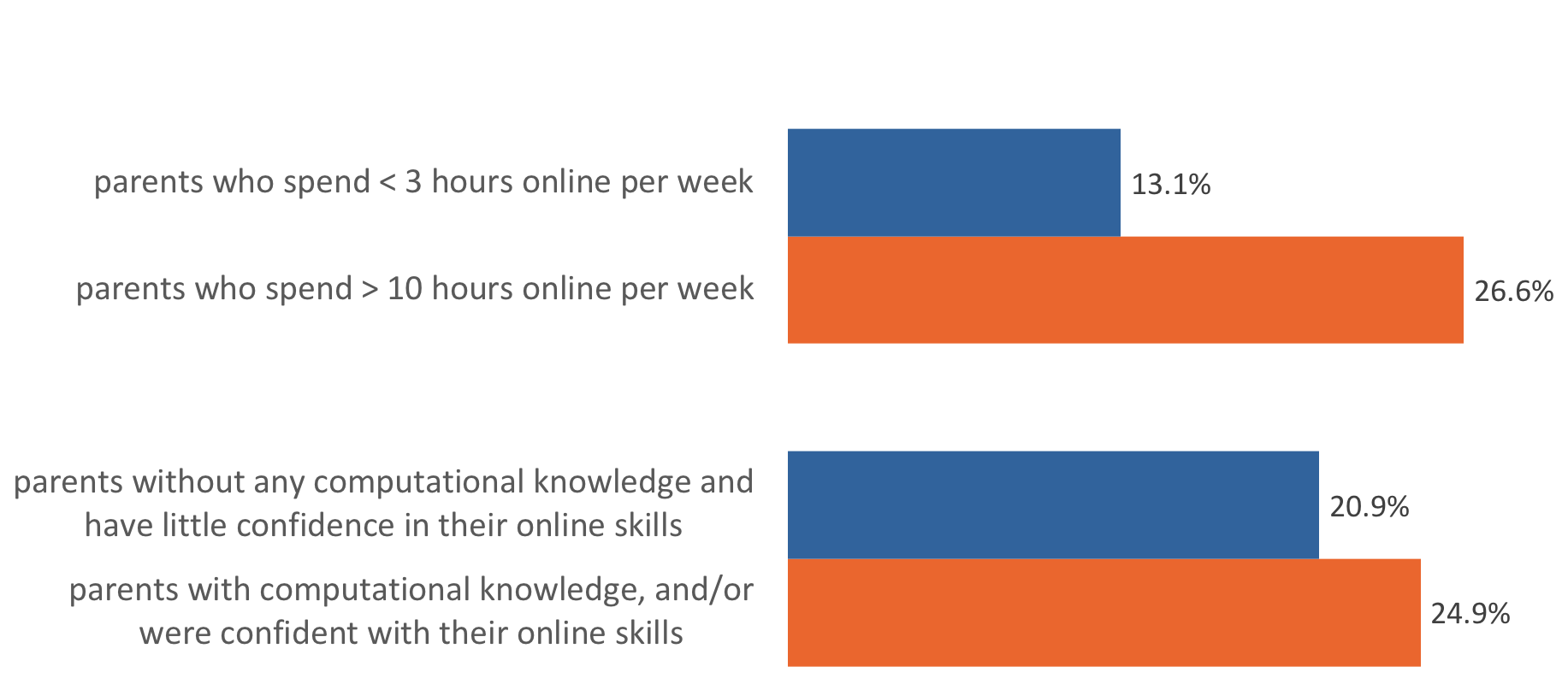}
\caption{More digitally experienced parents expressed more privacy concerns}
\label{fig:parconcerncorre}
\textbf{better labels for this figure 7}

\centering
\includegraphics[width=0.9\textwidth]{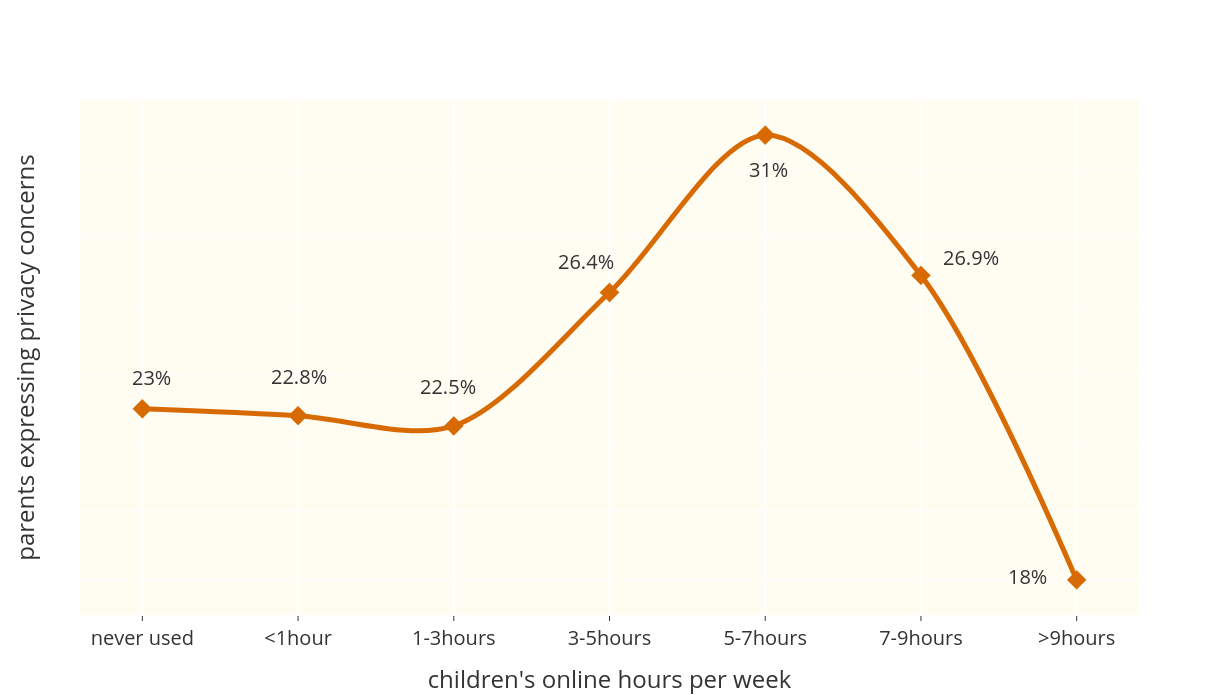}
\caption{children's online hours and their parents' awareness of privacy risks}
\label{fig:childconcerncorr}
\end{figure}

\begin{itemize}
    \item Parents who spend more time online themselves were more concerned with privacy issues around their child, nearly doubled those who spend less than 3 hours online per week.
    \item Parents with computational knowledge and/or claimed they are digitally-skilled also expressed more privacy concerns (Figure ~\ref{fig:parconcerncorre}).
\end{itemize}

We compared children's online hours per week and their parents' awareness of privacy risks, we found that: (Figure ~\ref{fig:childconcerncorr})
\begin{itemize}
    \item As children spend more time online, their parents' privacy concern grows.
    \item However, for the children who spend more than 9 hours online per week, only 18\% of those parents have expressed any kind of privacy concerns.
    \item This implies possibly a lack of understanding about the potential risks online (including privacy risks), and this lack of understanding led to less screen time control of children. 
\end{itemize}

\subsection{Parents' level of concern increased when told about potential implications }

\begin{figure}[h]
\centering
\includegraphics[width=0.7\textwidth]{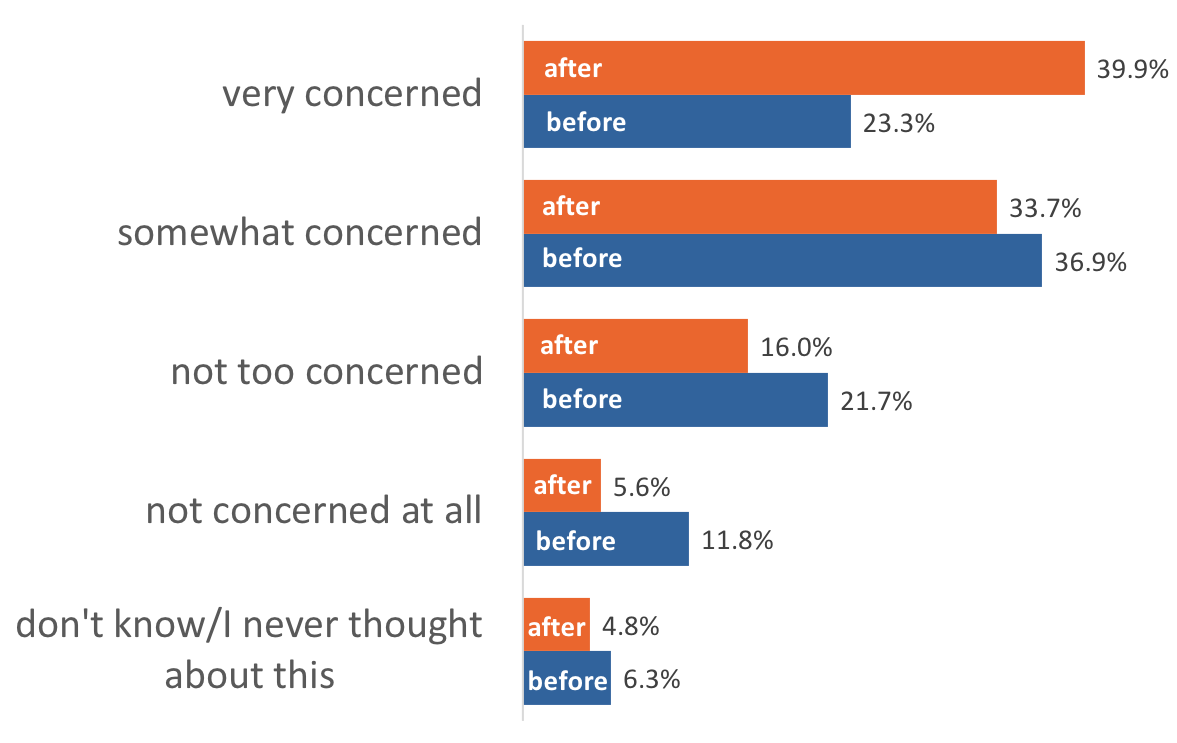}
\caption{Parents' level of concern regarding access to devices' camera, microphone and photo (Q24 vs Q25)}
\label{fig:24a}
\end{figure}

\begin{figure}[h]
\centering
\includegraphics[width=0.7\textwidth]{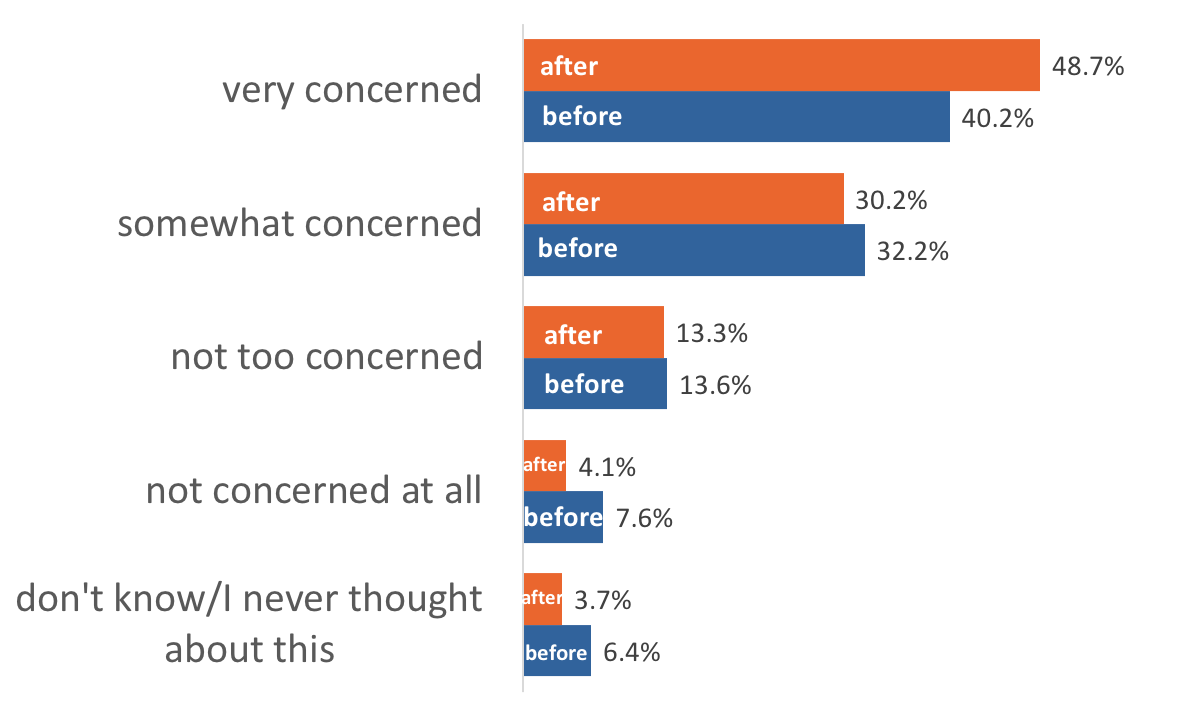}
\caption{Parents' level of concern regarding access to devices' location information (Q24 vs Q25)}
\label{fig:24b}
\end{figure}

\begin{figure}[h]
\centering
\includegraphics[width=0.7\textwidth]{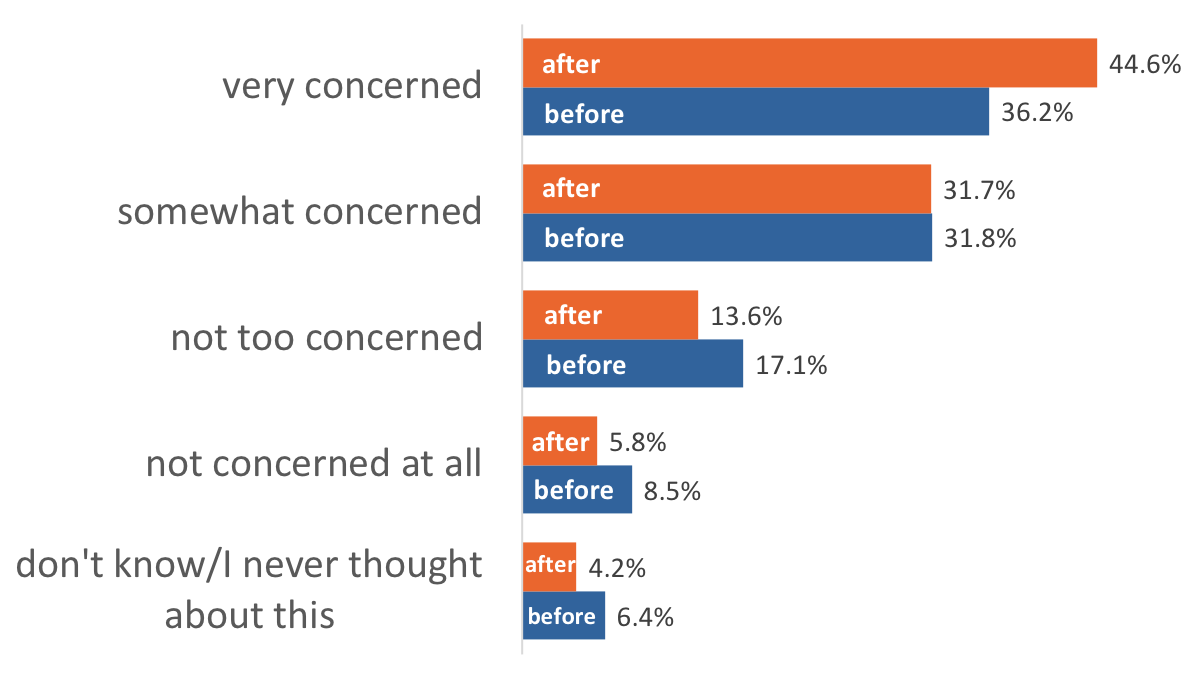}
\caption{Parents' level of concern regarding access to children' personal preference information (Q24 vs Q25)}
\label{fig:24c}
\end{figure}

While the majority of the parents said that they were concerned about apps accessing their children's camera, location information and personal preference information, parents' level of concern increased when they were presented with the possibility that these permissions might enable strangers to communicate with their child or share their photos, companies to infer sensitive information about their child and that companies might be able to generate personalised adverts based on children's preferences (Figure  ~\ref{fig:24a},~\ref{fig:24b},~\ref{fig:24c})

\begin{itemize}
    \item Out of the 3 scenarios, parents' level of concern increased most regarding access to devices' camera, microphone and photo.
    \item This implies possibly a lack of understanding about the consequences of having apps access to some critical information about a child's device. Strengthening parents' understanding about the implications of giving apps access to critical information about a device could potentially lead to more informed decisions.
\end{itemize}

Although parents' level of concern increased when told about potential implications of privacy risks, in comparison to UK parents, Chinese parents in our survey showed less concern in general. Only 39.9\% of them reported to be 'very concern', even after being told about the potential privacy risks of devices' camera, microphone and photo, this number is significantly less than the UK parents (75\%); Similar patterns were also found in the case of devices' location information, a much bigger proportion of UK parents were 'very concern' (69\%), comparing to Chinese parents (48.7\%).

\subsection{Parents adopted a range of means to safeguard their children online, however mostly through restrictive approaches} \label{parents_approach}

\begin{figure}[h]
\centering
\includegraphics[width=0.85\textwidth]{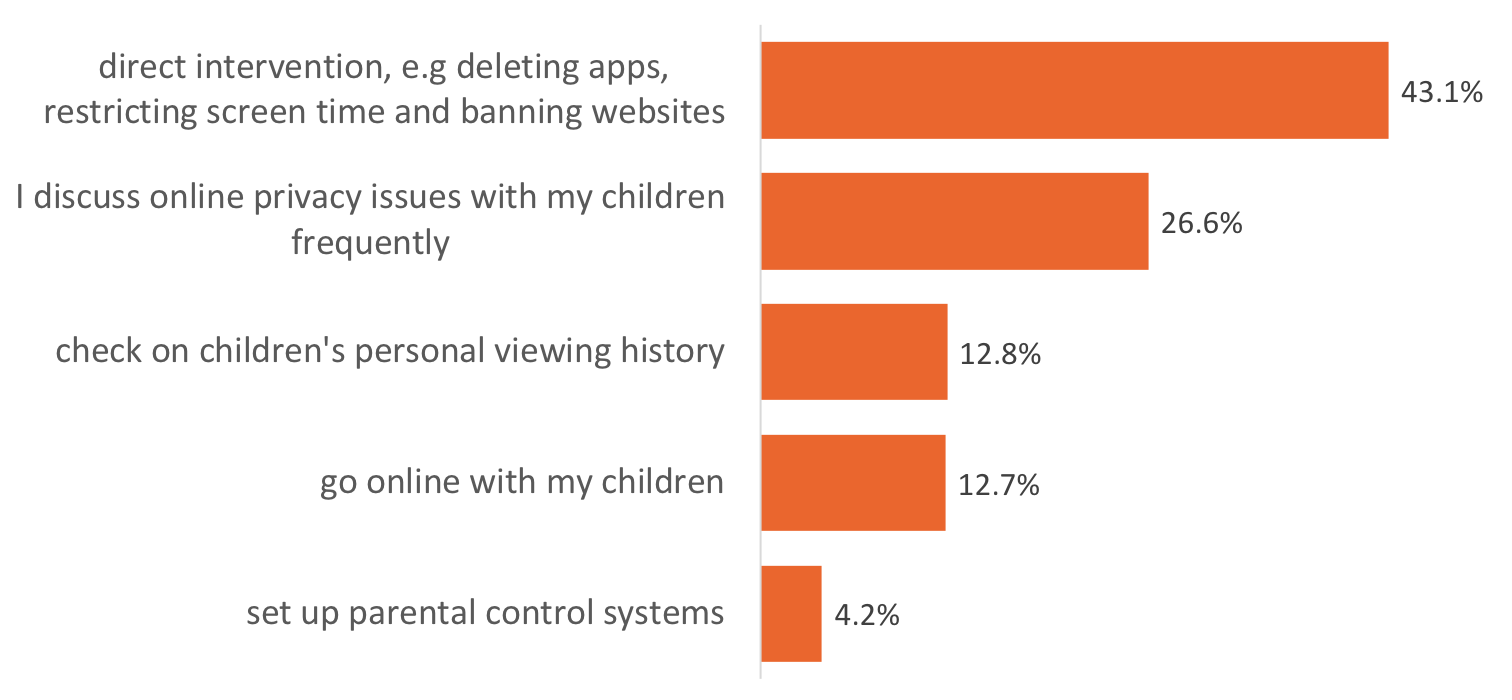}
\caption{Think about when you are safeguarding your child online, which one below is the current practice you have used most frequently? (Q19)}
\label{fig:q19}
\end{figure}

Parent respondents were asked to report their current practice when safeguarding their children online. While 59\% of them reported have used restrictive approaches, such as controlling children's screen time, monitoring their online activities and restricting the websites they could have access to; Only 26.6\% of parents said they had been discussing online privacy issues with their children. (Figure ~\ref{fig:q19})

Most parents (70\%) reported that they have taken some practices to safeguard their children online.
\begin{itemize}
    \item 29.7\% said they reviewed the app's privacy terms to make sure it's acceptable.
    \item 30.8\% said they checked the app's privacy settings.
    \item 35.6\% of them reported that they checked the app's age restriction and only used the ones that were under age limit for their children. However, this contradicts with what was reported in Q15: 8 out of the 10 most used apps had age restriction as at least 12+, which were not appropriate to used by 6-10 year-olds.
\end{itemize}

In the survey, parents were also asked to report how often they discuss online privacy issues with their children. Most parents (62\%) said that they have never or very seldom talked about these issues with their children; 53.4\% of them reported that their children have never learnt about online privacy risks at schools.

In comparison to UK parents, Chinese parents in our survey showed more tendency towards direct intervention. UK parents, according to the project 'Parenting for a digital future' \cite{livingstonedata}, when asking about the things they do in relation to children's internet use, the most frequent response from parents was to 'Make rules about how long or when your child is allowed to go online', followed by 'Talk to your child about what they do on the internet', 'Suggest ways that your child can use the internet safely' and 'Use parental controls or apps to block or monitor your child's access to some types of websites'. UK parents ranked direct intervention methods lower than methods involved communication with their children, while Chinese parents favored the previous one more. This links back to our findings that Chinese parents in survey were much more likely to uninstall their children's apps than UK parents, which may be related to the authoritarian parenting stlye more often observed in Chinese families.

\subsection{Communication with children about online risks is under good control by parents, as reported}
\begin{figure}[h]
\centering
\includegraphics[width=0.85\textwidth]{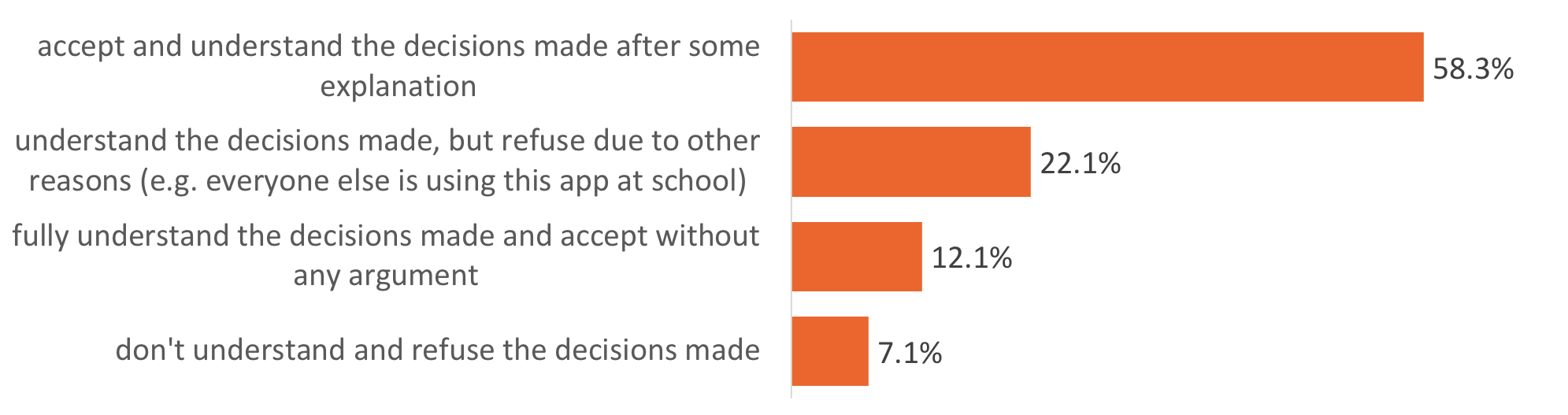}
\caption{When you are taking practices to safeguard you child online, what will be their most common responses? (Q20)}
\label{fig:q20}
\end{figure}

Parents did not struggle too much with managing children's emotions when safeguarding them online. Of the 69.6\% households who have reported that they had taken safeguarding practices for their children, most parents (58.3\%) said their children would understand parents' decisions after explaining the risks to them, 12.1\% of them even reported that their children would fully understand the decisions and accept without any argument. A smaller number of parents (22.1\%) reported their children would be able to understand them but refused due to some other reasons (e.g. peer pressure, everyone else is playing this game). Only 7.1\% reported that their children would be upset and refuse their parents' help. (Figure ~\ref{fig:q20}) 

Comparing to Chinese parents, UK parents struggled more with managing children's emotions when they were trying to remove inappropriate apps for children. 45\% of parents said their children would understand and accept the decisions made, which is a lot lesser than Chinese parents (70.4\%). Many UK parents said that their children (48\%) were very upset and could not understand their parents' decisions, while only 7.1\% of Chinese parents reported the same. At the same time, however, Chinese parents were found to 'talk less' with their children about their online safety (Section \ref{parents_approach}), this again relates back to the authoritarian parenting style more often observed in Chinese parents.

\begin{figure}[h]
\centering
\includegraphics[width=0.8\textwidth]{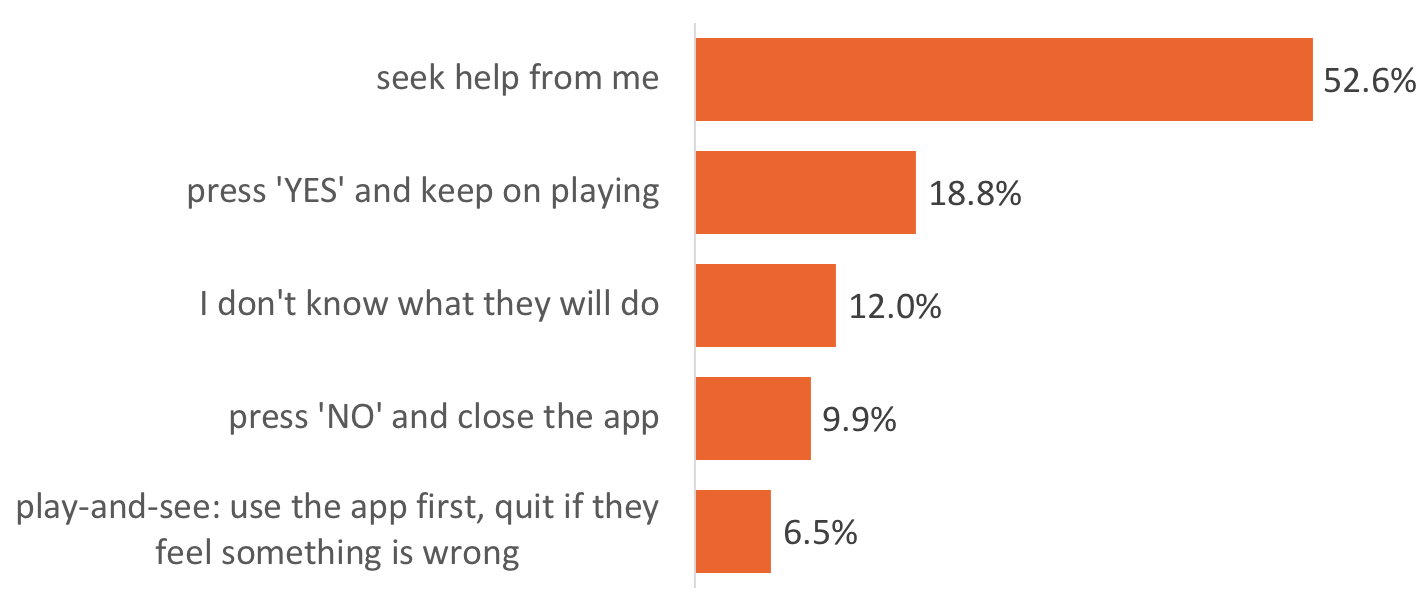}
\caption{When your child's app is asking consent to the above information (e.g. camera, microphone, location information, personal preferences), what do you think your children would do? (Q26)}
\label{fig:q26}
\end{figure}

In the survey, we also asked parents to think about what their children will do if they were asked to expose their personal information online or give consent to apps (Figure ~\ref{fig:q26}).

\begin{itemize}
    \item 12\% of parents said they don't know what will their child do. 18.8\% said their children will not pay attention to the consent notice, hit 'YES' and keep on playing.
    \item The majority of parents (52.6\%) said that their children would seek help from them. 6.5\% thought their children would not be able to fully recognise the potential risks therefore adopt the 'play-and-see' strategy. 
    \item These reactions were also commonly found in our previous focus group studies with UK children \cite{10.1145/3290605.3300336}. These observations are in support of our findings, such that parental involvement would lead to improved learning outcomes of children. However, parents themselves might have not been well supported in dealing with challenges related to facilitating their children's use of digital technologies.
\end{itemize}

\subsection{Parents need most help on knowing the exact things their children have been doing online}

\begin{figure}[h]
\centering
\includegraphics[width=1\textwidth]{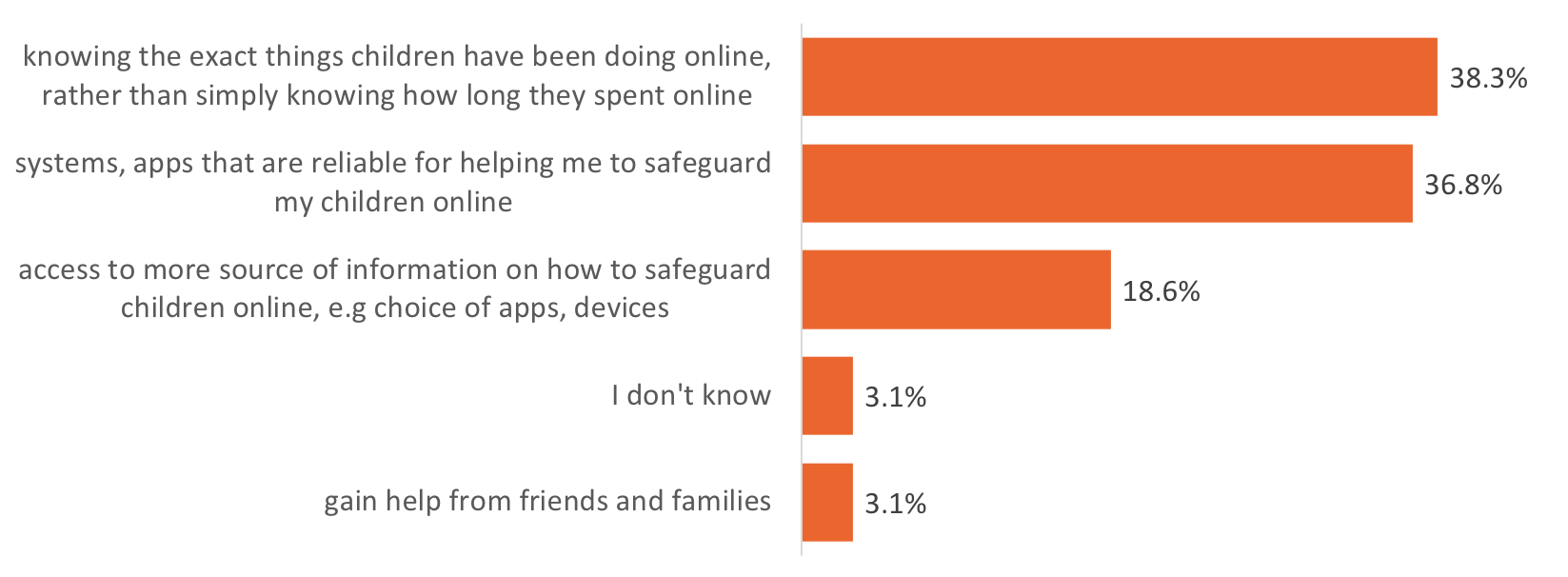}
\caption{Which aspect do you think you need most help with? (Q29)}
\label{fig:q29}
\end{figure}

When asked about what aspect do they need most help on, most parents agreed on that they want to learn more about the exact things their children have been doing online, rather than just knowing how long they have spent on their digital devices, and a  large amount of parents wanted these helps to be delivered through apps or systems that were reliable and specifically designed for them (Figure ~\ref{fig:q29}). 

This have raised important implications that parents are in need of such apps/systems/knowledge sources that best meet their needs and help them with their safeguarding practices. Most of the time they rely on self-guided online searches, rather than being informed by systematic, comprehensive and reliable resources \cite{livingstone2018parentsa}. Future tool development should consider both scaffolding children’s knowledge acquisition and facilitating the active involvement of the parents.

%% file: conclusion.tex
\section{Conclusion}

In this report we presented the responses of an online survey towards understanding what online threats and especially those privacy related ones, children ages 6-10 and their parents were aware of during the interaction with digital devices. We found that digital technologies have taken an important role in the daily life of Chinese families, and that although our parent respondents had a reasonable awareness of the potential risks online, like inappropriate content or excessive screen time, they largely have little awareness of the online privacy risks. Parents have good knowledge and have applied a range of different means to safeguard their children online, however, this are typically through restrictive approaches and they struggled to discuss these issues with their children, partly because they need help on strengthening their technical competence, and they need to be better informed on how their choices and decisions would impact on their children's online privacy safety.

Parents' concerns about content appropriateness and age restrictions have not always led to consistent choice of apps for children. This indicates a need for further understanding of why this is happening and how we may provide better support for the parents. We also observed difference in app usage between UK and Chinese children: 'social network/media' apps like WeChat and Tiktok, 'education' apps were used a lot by Chinese children, which is quite different to UK children, who use 'online video platforms' and 'game' apps more often. At the same time, we also found that authoritarian parenting style was more often found in Chinese families, Chinese parents adopted a range of means to safeguard their children online, however mostly through restrictive approaches.

We also found that Chinese parents are in need of better support on more reliable and accessible apps and tools that could help them with their safeguarding practices and to help them in transmitting the essential knowledge and skills to young children, who are at the frontier of risks. 

Based on these findings, we recommend:
\begin{itemize}
    \item Raising the general awareness of online privacy risks for both parents and children, and facilitating these discussions with young children.
    \item Encouraging parents to discuss online safety issues with their children, which provides the necessary scaffolding process for children's learning. This would require resource development that help parents improve their digital skills and digital confidence.
    \item Tool and resource developments that focus on facilitating skill and knowledge building for both parents and their young children that enables parents to learn more about their children's online activities, and encouraging an active co-learning experience.

\end{itemize}